\documentclass[12pt]{article}
\usepackage[dvips]{graphicx}
\usepackage{amsmath,amsthm,amssymb,amscd,mathtools}
\usepackage[noblocks]{authblk}
\usepackage{mathrsfs}
\usepackage{threeparttable}
\usepackage{natbib}
\newtheorem{theorem}{\bf \large Theorem}
\allowdisplaybreaks
\def\bm#1{\hbox{\boldmath  $#1$\unboldmath}{}}
\usepackage{titling}
\usepackage{graphicx}
\usepackage{threeparttable}
\usepackage{dcolumn}
\usepackage{multirow}
\usepackage{booktabs}
\usepackage{color}
\usepackage{epsfig}
\usepackage{rotating}
\usepackage{algorithm}
\usepackage{geometry}
\usepackage{subfig}
\usepackage{caption}
\usepackage{ulem}

\makeatletter
\newenvironment{breakablealgorithm}
{
	\begin{center}
		\refstepcounter{algorithm}
		\hrule height.8pt depth0pt \kern2pt
		\renewcommand{\caption}[2][\relax]{
			{\raggedright\textbf{\ALG@name~\thealgorithm} ##2\par}%
			\ifx\relax##1\relax 
			\addcontentsline{loa}{algorithm}{\protect\numberline{\thealgorithm}##2}%
			\else 
			\addcontentsline{loa}{algorithm}{\protect\numberline{\thealgorithm}##1}%
			\fi
			\kern2pt\hrule\kern2pt
		}
	}{
		\kern2pt\hrule\relax
	\end{center}
}
\makeatother

\begin{document}

\title{\large Exact Inference for Common Odds Ratio in Meta-Analysis with Zero-Total-Event Studies}

\vspace{-20mm}

\author{\normalsize Xiaolin Chen$^{a}$, Jerry Cheng$^b$, Lu Tian$^c$ and Minge Xie$^d$ \\
\normalsize $^a$School of Statistics and Data Science, Qufu Normal University	\\
\normalsize $^b$Department of Computer Science, New York Institute of Technology \\
\normalsize $^c$ School of Medicine, Stanford University
\\
\normalsize $^d$Department of Statistics, Rutgers University}

\date{}

\maketitle

\vspace{-16mm}

\begin{abstract}
Stemming from the high profile publication of Nissen and Wolski (2007) and subsequent discussions with divergent views on how to handle observed zero-total-event studies, defined to be studies which observe zero events in both treatment and control arms, 
the research  
topic concerning the common odds ratio model with
zero-total-event studies remains to be an unresolved problem in meta-analysis. 
In this article, we address 
this problem 
by proposing a novel repro samples method to handle zero-total-event studies and make inference for the parameter of common odds ratio. 
The development  explicitly accounts for sampling scheme and
does not rely on large sample approximation.  
It is theoretically
justified with a guaranteed finite sample performance. 
The empirical performance of the proposed method is demonstrated through simulation studies. It shows that the proposed confidence set achieves the desired empirical coverage rate and also that the zero-total-event studies contains information and impacts the inference for the common odds ratio. The proposed method is also applied to combine information 
in the Nissen and Wolski study.  
\end{abstract}

\noindent \textbf{Keywords:} Exact confidence interval; Meta-analysis; Odds ratio; Repro samples; Zero-total-event studies

\section{Introduction}

Meta-analysis methodology developed for synthesizing information across multiple independent (but comparative) sources has a long history and remains to be a popular research topic in statistics \citep{Breslow1981,Normand1999,SuttonHiggins2008,Xieetal2011,Cooper2019}. It is particular useful for the settings
where a single study is inadequate for drawing a reliable conclusion and conclusions can often be strengthened by aggregating information from all studies of the same or a similar kind.  Meta-analysis approaches have become a widely used tool in many fields, such as biomedical research, pathology, library and information science, education and so on. 
One of the research topics in meta-analysis that remain open is how to handle an observed zero-total-event study that is defined to be a study which observes zero events in both treatment and control arms \citep[cf.,][]{FinkelsteinLevin2012, LiuLiuXie2014, YangLiuWangXie2016}. 
This problem has long been debated since the high profile publication by \citet{NissenWolski2007}, as there are divergent but inclusive views on how to the handle zero-total-event studies \citep{FinkelsteinLevin2012, Xieetal2018}. 
In this article, we revisit this problem 
and propose a novel exact meta-analysis procedure to handle zero-total-event studies. 

Our research is motivated by the exact study of \citet{NissenWolski2007}
on drug safety evaluation of the use of diabetic drug Avandia. 
In \citet{NissenWolski2007}, the authors collected data from 48 clinical studies, and conduct a meta-analysis to assess whether Avandia significantly increases the risk of myocardial infraction and death from cardiovascular diseases. Most of these studies reported zero or a very small number of  events in one or both of treatment and control groups. 
\citet{NissenWolski2007} used Peto's method to combine information across all studies, which effectively discarded more than half of 48 studies in the analysis of endpoint cardiovascular death (25 out of the total 48 studies are zero-total-event studies). This practice was challenged by \cite{diamond2007uncertain}, initiated a hot debate in the community with diverging views on how to handle observed zero-total-event studies in general.  
The key difficulties are that $0/0$ has no mathematical definition and also that most of the existing meta-analysis methods rely on normality or large sample justifications and therefore are not suited for analysis of zero-total-event studies. Indeed, 
as stated in \cite{Xieetal2018}, with the probabilities of both treatment and control events $(\pi_{0i}, \pi_{1i})$ not equal to
$0$ (even though very small), the probability of observing a zero-total-event study is $0$
when the number of patients in both treatment arms $n_i \to \infty$ and $m_i \to \infty$. Thus, when a zero-total-event study
is observed, it is an indication that the sample sizes are not
large enough for this particular underlying set.
Until today, the statistical inference problem at the center of this debate is still open and unanswered \citep{FinkelsteinLevin2012,Xieetal2018}.

Consider a typical setting of $K$ independent clinical trials (control vs treatment): $X_i \sim$ Binomial$(n_i,$ $\pi_{0i})$ and $Y_i$ $\sim$ Binomial$(m_i, \pi_{1i})$,  $i = 1, \dots, K$. 
We can often express  the sample data in $K$ $2 \times 2$ tables:
\begin{table}
\caption{$2\times2$ Clinical Study with Control and Treatment}
\begin{gather}
\label{2by2_i}
\resizebox{.90\textwidth}{!}{$\begin{array}{r|cc|c}
 & \hbox{Yes}  &  \hbox{No} &   \\
\hline
\hbox{Control} & X_i  &  n_i - X_i  & n_i  \\
 \hbox{Treatment} & Y_i  &  m_i - Y_i & m_i  \\
\hline
\hbox{Total}  & X_i+ Y_i  & (n_i + m_i ) - (X_i+ Y_i) & n_i + m_i \\
\end{array}
\qquad \hbox{for \quad $i = 1 ,2 , \ldots, K$,} $}\nonumber
\end{gather}
\end{table}
where $X_i$ and $Y_i$ are the  numbers of events in the control and treatment arms of the $i^{th}$ trial. Often $(\pi_{0i},\pi_{1i})$ is reparameterized to $(\theta_i, \eta_{i})$, with the log odds ratio
{\small $\theta_i = \log \big({\pi_{1i} \over 1-\pi_{1i}}\big/{\pi_{0i}\over 1-\pi_{0i}}\big)$} and {\small $\eta_{i}=$ $\log\big\{\big(\frac{\pi_{1i}}{1-\pi_{1i}}\big)$ $\big(\frac{\pi_{0i}}{1-\pi_{0i}}\big)\big\}$}. A classical common odds ratio model assumes $\theta_1$ $\equiv$ $\ldots$ $\equiv$ $\theta_K$ $=$ $\theta$, but the rates $(\pi_{0i},\pi_{1i})$ allow to be different from one study to another; cf.,  
\cite{Breslow1981,CoxSnell1989, NissenWolski2007, FinkelsteinLevin2012, Tianetal2009}, among others.
In {{\it rare event studies}, both $\pi_{0i} > 0$ and $\pi_{1i} > 0$ but are very small}. In this case, the observed data, say $\{x_i^{obs}, y_i^{obs}\}$, can often be $0$ or very small numbers 
($n_i$ and $m_i$ can be large typically in thousands). The studies with observed data $x_i^{obs} = y_i^{obs} = 0$ are referred  to as {\it zero-total-event studies}  in the literature  \citep[cf.,][]{FinkelsteinLevin2012, LiuLiuXie2014}.  
In this article, we focus on the inference problem of $\theta$, or more specifically, constructing a finite-sample performance guaranteed level-$\alpha$ confidence interval for $\theta$ in meta-analysis while incorporating potentially many zero-total-event studies.

The analysis of rare event data, in particular incorporating zero-total-event studies in a meta-analysis, raises specific statistical challenges and has been intensely studied \citep{Sweetingetal2004,Bradburnetal2007,FinkelsteinLevin2012,Tianetal2009,Caital2010,Bhaumiketal2012,LiuLiuXie2014,YangLiuWangXie2016}.
Most commonly used meta-analysis methods rely on the asymptotic distribution of the combined estimator to make inference. For instance, the widely used inverse-variance weighted method combines point estimators from individual studies, assuming that the distributions of all the estimators can be well approximated by normal distributions. The classical Mantel-Haenszel \citet{MantelHaenszel1959} and Peto methods \citet{Yusufetal1985} also rely on the normal approximation to the distribution of the combined estimator. However, the normal approximations are ill-suited  for rare events data and results for rare events data in practice often 
yield an unacceptably low coverage probability \citep{Bradburnetal2007,Tianetal2009}. In addition, the commonly practiced ``continuity correction'' (i.e. adding 0.5 or 0.1 to zero cells) is shown with compelling evidence to have undesirable impact on inference outcomes \citep{Sweetingetal2004,Bradburnetal2007}. Conditional likelihood inference methods have also been proposed for meta-analysis of $2\times2$ tables \citep[e.g.,][]{CoxSnell1989}. In particular, one can make inference relying on a conditional likelihood function and finite sample Fisher exact test, for which computing algorithms and small sample approximations are developed \citep{Mehtaetal1985, Davison1988}. Under the conditional inference framework, the conditional likelihood function of a zero-total-event study is constant, and thus the study does not contribute to the inference.
However, based on the likelihood principle \citep{BergerWolpert1988}, \cite{Xieetal2018} showed that the conditional likelihood, although maintaining test size, loses power (compared to the full likelihood method) and Fisher exact test is not particularly suited for analysis of  zero-total-event clinical trials, a conclusion also reached independently in \cite{FinkelsteinLevin2012}. Bayesian methods have also been experienced to analyze zero-total-event studies, 
in which zero-total-event studies typically
contribute to the meta-analysis inference. Since the use of priors imposes an additional model assumption and rare events data are very sensitive to the prior choices, it is argued in the field that a Bayesian approach ``may raise more questions than they settle'' (cf, \cite{FinkelsteinLevin2012}). In recent years, several finite sample methods are proposed for rare events data but for different inference problems. For instance, \cite{Tianetal2009} proposes an exact method for meta-analysis of risk difference $p_{1k}-p_{0k}$. Although \cite{Tianetal2009} does not use large sample approximations, it is on risk difference and cannot handle the parameter of odds ratio. \cite{YangLiuWangXie2016} reviews exact meta-analysis methods with a focus on rare events and shows that the method by \cite{Tianetal2009} is a spacial case of \cite{Xieetal2011}. \cite{Caital2010} suggests to use a Poisson model to analyze the rare event $2\times2$ tables. The approach avoids the difficult question of $0/0$, but by changing the distribution assumption it also changes the original inference target in the two~binomial $2\times2$ tables. 

Despite all the efforts, it remains an open and unanswered inference problem in statistics on how to handle the zero-total-event studies in analysis of the common odds ratio \citep{FinkelsteinLevin2012,Xieetal2018}. 
The debate on zero-total-event studies are centered on two questions: (a) Does a zero-total-event study  possess any information concerning the parameter of common odds ratio? (b) If it does, how can we effectively incorporate zero-total-event studies in meta-analysis?
In \citet{Xieetal2018}, the authors showed that zero-total-event studies indeed possess information about the parameter common odds ratio in meta-analysis. In the current article, we provide a solution to the second question on how to effectively include zero-total-event studies to help make an effective inference on the common $\theta$ in meta-analysis. 

Our solution is developed based on a newly developed inferential framework called {\it repro samples method} \citep{XieWang2022}.  
The repro samples method uses a simple yet fundamental 
idea: Study the performance of artificial samples that are generated by mimicking the sampling mechanism 
of the observed data; the 
artificial samples 
are then used to help 
quantify the uncertainty in estimation of model and parameters.  
The repro samples development is deeply rooted and grown from prior developments of artificial-sample-based inference procedures across Bayesian, frequentist and fiducial paradigms (i.e., approximate Bayesian computing, Bootstrap, generalized fiducial inference and inferential model; See further discussions in \citealp{XieWang2022}). 
It does not need to rely on likelihood functions or large sample theories, and it is especially effective for difficult inference problems in
which regularity conditions and thus regular inference approaches do not apply. \cite{XieWang2022} and \cite{WangXieZhang2022} used the repro samples framework to address two open inference questions in statistics concerning (a) Gaussian mixture and (b) high dimensional regression models, where the authors successfully provided finite-sample confidence set for discrete unknown parameters (i.e., unknown number of components in the mixture model and unknown sparse model in the high dimensional model) along with joint confidence sets for the unknown discrete and also the remaining model parameters. In our current paper, our problem does not involve any discrete parameters, however we can still use some of the key techniques in the repro samples framework to develop a novel methodology with finite sample supporting theories to address the highly non-trivial inference problem concerning zero-total-event studies.

The rest of this article is organized as follows. Section 2 introduces the repro samples method and our proposed inference procedure. Section 3 provides extensive simulation studies to examine the performance of proposed method and compare it with the popular Mantel-Haenszel and Peto methods. A new analysis of the Avandia data in \cite{NissenWolski2007} using the proposed repro samples method is  provided in Section 4. A brief summary and discussion is given in Section~5.

\section{Repro Samples Method for Meta-analysis of $2 \times 2 $ Tables}
Since the repro samples method is relatively new, we first provide in Section 2.1 a brief description of
the method, based on which we provide our new development tailored to zero-total-event studies in Sections 2.2 and 2.3. 

\subsection{Notations, terminologies and brief review of repro samples method}

Suppose the sample data 
${Y}$ $\in {\cal Y}$
are generated from 
an {\it algorithmic model}:
\begin{equation}\label{eq:1}
{Y} = G({\theta}, U), 
\end{equation} 
where $G(\cdot, \cdot)$ is a known mapping from $\Theta \times {\cal U} \mapsto {\cal Y}$, $\theta \in \Theta$ is model parameter and ${U} = (U_1, \ldots U_r)^\top \in {\cal U}
\subset R^r$, $r>0$, is a random vector whose distribution is known or can be simulated.
Thus, given~${\theta} \in \Theta$, we know how to simulate data $Y$ from (\ref{eq:1}).
In fact, this is the only assumption needed in the repro samples development. The model $G(\cdot, \cdot)$ can be very complicated in either an explicit or in-explicit form, including complex examples such as differential equations or generative neural networks. As long as we can generate $Y$ for a given $\theta$, we can apply the method.  
Denote by the observed data ${y}^{obs} = G({\theta}^{(o)}, {{u}}^{rel})$, where ${\theta}^{(o)} \in \Theta$ is the true value 
and ${{u}}^{rel}$ 
the corresponding (unknown) realization of ${{U}}$.

Let $T(\cdot, \cdot)$ be a mapping function from ${\cal U} \times \Theta \to$ ${\cal T} \subseteq R^{q}$, for some $q \leq n$. Also, for each given ${\theta}$, let $B_{\alpha}({\theta})$ be a Borel set such that 
\begin{equation}
    \label{eq:B}
    \mathbb{P} \left\{T({U}, {\theta})  \in B_{ \alpha}({\theta}) \right\}\ge \alpha, \quad 0 < \alpha < 1. 
\end{equation}

\noindent
The function $T$ is refeered to as a {\it nuclear mapping} function. A repro samples method 
constructs a subset in $\Theta$: 
\begin{equation}
\label{eq:G0}
\Gamma_{\alpha}({y}^{obs}) = \big\{{\theta}: \exists \, {u}^* \in {\cal U} \mbox{ such that }  {y}^{obs} = 
G({{\theta}}, {u}^*),
\,  
T({u}^*, {\theta})  \in B_{\alpha}({\theta})  \big\} \subset \Theta, 
\end{equation}
In another words, for a potential value ${\theta}$, if there exists a ${u}^*$ such that the artificial sample ${{y}}^* = G( {{\theta}}, {{u}}^*)$ matches ${y}^{obs}$ (i.e., ${y}^* = {y}^{obs}$) and $T({u}^*, {\theta})  \in B_{\alpha}({\theta})$, then we keep this ${\theta}$ in the set. Since ${y}^{obs} = G({{\theta}}^{(o)},$ ${u}^{rel})$, if 
$T({u}^{rel}, {\theta}^{(o)})  \in B_{\alpha}({\theta}^{(o)})$, then ${\theta}^{(o)} \in \Gamma_{\alpha}({y}^{obs})$. 
Similarly,
under model ${Y} = G({{\theta}}^{(o)}, {U})$, if $T({U}, {\theta}^{(o)}) \in B_\alpha({\theta}^{(o)})$, then ${\theta}^{(o)} \in \Gamma_{\alpha}({Y})$. Thus, by construction, 
$\mathbb{P} \big\{{\theta}^{(o)} \in \Gamma_\alpha({Y}) \big\} \ge \mathbb{P} \big\{T({U}, {\theta}^{(o)}) \in B_\alpha(\theta^{(o)}) \big\}  \ge \alpha.
$ 
This proves that $\Gamma_{\alpha}({y}^{obs})$ 
is a level-$\alpha$ confidence 
set for ${\theta}_0$. This development is likelihood-free and does not need to rely on any large sample theories. 

The repro samples development  utilizes the ideas of  {\it inversion} and {\it matching of artificial and observed samples}. Let's illustrate the development using a very simple toy example of $Y \sim N(\theta,1)$. In the form of (\ref{eq:1}), $Y = \theta + U$, where $U \sim N(0,1)$. Suppose the true underlying parameter value is $\theta^{(o)} = 1.35$ and the realization is $u^{rel} = 1.06$, giving us a single observed data point $y^{obs} = 2.41$.  We only know $y^{obs} = 2.41$ and $u^{rel}$ is a realization from $N(0,1)$ but we do not know its value $1.06$. We would like to make an inference for $\theta^{(o)}$. Let $T(U, \theta) = U$, then 
the level-$95\%$ Borel set in (\ref{eq:B}) is the interval $(-1.96, 1.96)$. By (\ref{eq:G0}), we keep and only keep those potential $\theta$ values that can reproduce $y^{obs} = 2.41$ by setting (matching) $\theta + u^* = 2.41$ with a (potential) realized error $u^* \in (-1.96, 1.96)$. This method of getting the set of $\theta$'s is essentially an inversion procedure and the method leads us to a level-$95\%$ confidence set $(0.45, 4.37)$, which is exactly the same best possible level $95\%$ confidence interval when observing a single data point $y^{obs} = 2.41$ using the classical frequentist method. 

The repo samples method does not need to involve the likelihood function and has a finite sample performance guarantee. \cite{XieWang2022} also showed that the repro methods is more general and flexible and subsumes the Neyman-Pearson framework as a special case. 
By using the repro samples development in our current paper on meta-analysis of $2 \times 2$ tables, we ask, for a potential value of the common log odds ratio parameter $\theta$ and a given confidence level $\alpha$, whether the $\theta$ value can be potentially be used to generate an artificial data set that match the observed studies. If it does, we keep the $\theta$ value in our level-$\alpha$ confidence set. One complication is that there are also nuisance parameters $\bm{\eta} = (\eta_1, \ldots, \eta_K)^T$. We provide our detailed development in Section 2.2.

\subsection{Repro samples method and finite-sample confidence set for the common odds ratio in $2\times2$ tables}

For the common odds ratio model in the $2 \times 2$ tables. We have $\pi_{0,i} = e^{(\theta + \eta_{i})/2}$ and $\pi_{1,i} = e^{(\theta - \eta_{i})/2}$, for $i = 1, \ldots, K$. 
We write $\bm{X}=(X_{1},\cdots,X_{K})^{T}$ and
$\bm{Y}=(Y_{1},\cdots,Y_{K})^{T}$. 
In the form of (\ref{eq:1}), the pair of binomial models $X_i \sim$ Binomial$(n_i,$ $\pi_{0i})$ and $Y_i$ $\sim$ Binomial$(m_i, \pi_{1i})$,  $i = 1, \ldots, K$, can be re-expressed~as
\begin{equation}
    \label{eq:G2by2}
    X_i = \sum_{j = 1}^{n_i} I\{ U_{ij} \leq e^{(\theta + \eta_{i})/2} \}\hbox{ and } Y_i = \sum_{j = 1}^{m_i} I\{V_{ij} \leq e^{(\theta - \eta_{i})/2} \}, \quad \hbox{for $i = 1, \dots, K$,}
\end{equation}
where $U_{ij}$ and $V_{ij}$ are iid $U(0,1)$ distributed random variables, for $j = 1, \ldots, n_i$ or $m_i$, $i = 1, \dots, K.$
We observe 
 $\bm{x}^{obs}=(x^{obs}_{1}, \ldots,x^{obs}_{K})^{T}$ and
$\bm{y}^{obs}=(y^{obs}_{1},\ldots,y^{obs}_{K})^{T}$,
$x_i^{obs} = \sum_{j = 1}^{n_i} I\{ u_{ij}^{rel} \leq e^{(\theta^{(o)} - \eta_{i}^{(o)})/2} \}$ and $y_i^{obs} = \sum_{j = 1}^{n_i} I\{ v_{ij}^{rel} \leq e^{(\theta^{(o)} - \eta_{i}^{(o)})/2} \}$, where  
$\theta^{(o)}$ 
and $\bm{\eta}^{(o)} =(\eta_{1}^{(o)},
\ldots,\eta_{K}^{(o)})^{T}$ 
are the true parameter values and $\bm{u}_i^{rel} = (u_{i1}^{rel}, \ldots, u_{im_i}^{rel})^T$
$\bm{v}_i^{rel} = (v_{i1}^{rel}, \ldots, v_{im_i}^{rel})^T$ are the corresponding realized random vectors that generated $\bm{x}^{obs}$ and $\bm{y}^{obs}$, respectively. 
The number of tables $K$ and each table's $(n_i, m_i)$ are given (not need to go to infinity). 
Among the $K$ tables, we allow many zero-total-event studies with $ x_i^{obs} = y_i^{obs} = 0$, but assume that at least one of $x_i^{obs} \not = 0$ and one of $y_i^{obs} \not = 0$. Our goal is to use a repro sample method to construct a performance guaranteed level-$\alpha$ confidence interval for the common log odds ratio parameter $\theta^{(o)}$ while taking care of the remaining $K$ nuisance model parameters $\eta_{i}$, $ i = 1, \ldots, K$.

Mantel-Haenszel statistic is a commonly used estimator of common log odds ratio,  
$$W_{MH}(\bm{X},\bm{Y})=
\log\Bigg(\sum_{k=1}^{K}R_{k}/\sum_{k=1}^{K}S_{k}\Bigg) 
$$ 
 where 
$R_{k}=X_{k}(m_{k}-Y_{k})/(m_{k}+n_{k})$ and $S_{k}=Y_{k}(n_{k}-X_{k})/(m_{k}+n_{k})$. 
To make inference, the Mantel-Haenszel method uses the large sample theorems by which 
\begin{equation}
  \label{eq:W-Theta}
W(\bm{X},\bm{Y};\theta) = W_{MH}(\bm{X},\bm{Y}) - \theta
\end{equation}
is normally distributed as both ${n_i} \to \infty$ and ${m_i} \to \infty$, for all $i = 1, \ldots, K$ \citep{Hauck1979,Breslow1981}. 
In rare events studies especially those contain zero-total-event studies, the large sample theorems do not apply, so a use of Mantel-Haenszel method is not theoretically justified for zero-total-event studies. However, due to its simplicity and 
good empirical performance especially in large sample situations, we use  $W(\bm{X},\bm{Y};\theta)$ in (\ref{eq:W-Theta}) to help develop the nuclear mapping function in our repro samples method to obtain a performance guaranteed finite sample confidence interval for~$\theta$.

For the sample data generated with parameter values $(\theta, \bm{\eta}^T)$, 
$X_{i} = \sum_{j=1}^{n_i} I\{U_{ij} \leq $ $e^{(\eta_{i}+\theta)/2}\}$ and 
$Y_{i} = \sum_{j=1}^{m_i} I{\{V_{ij} \leq e^{(\eta_{k}-\theta)/2}\}}$, 
the distributions of 
$W(\bm{X},\bm{Y};\theta)$ depends on the $K$ nuisance parameters $\bm{\eta} = (\eta_{1}, \ldots, \eta_{K})^T$. We use a profile approach to control the impact of the nuisance parameters $\bm \eta$.  Specifically, 
let $\widetilde X_{i} = \sum_{j=1}^{n_i} I{\{U_{ij}' \leq e^{(\widetilde \eta_{i}+\theta)/2}\}}$ and $\widetilde Y_{k} = \sum_{j=1}^{m_i} I{\{V_{ij}' \leq e^{(\widetilde \eta_{i}-\theta)/2}\}}$,
where $U_{ij}'$ and $V_{ij}'$ are iid $U(0,1)$ distributed random variables. We define, for $t \ge 0$, 
\begin{equation}
\gamma_{(\theta, \widetilde \eta)}\{t\} = 
\mathbf{P}
\left\{ \big|W(\widetilde{\bm{X}},
\widetilde{\bm{Y}};\theta)\big| < t
\right\}.
\end{equation}
In the special case with $\widetilde{\bm{\eta}} = \bm{\eta}$, we have $\gamma_{(\theta, \eta)}\{|W(\bm{X},\bm{Y};\theta)|\} \sim U(0,1)$. 
In particular, we can show that $1 - \gamma_{(\theta, \widetilde \eta)}\left\{|W(\bm{x},\bm{y};\theta)|\right\}  = \mathbf{P}
\big\{ \big|W(\widetilde{\bm{X}},\widetilde{\bm{Y}};\theta)\big| \geq  \big|W(\bm{x},\bm{y};\theta)\big|
\big\}$ is the $p$-value to reject the null hypothesis $H_0:$ a sample dataset $(\bm{x},\bm{y})$ is generated from $(\theta, \tilde {\bm\eta}^T)$, when in fact the a sample dataset $(\bm{x},\bm{y})$ is generated from $(\theta, {\bm\eta}^T)$.

Following the profile method proposed in \cite{XieWang2022}, we define our nuclear mapping function as
\begin{equation}
\label{eq:NT} T(\bm{X},\bm{Y};\theta) = \min_{\widetilde \eta \, \in \, \mathbf{R}^K} \gamma_{(\theta, \widetilde \eta^T)}\left\{|W(\bm{X},\bm{Y};\theta)|\right\}
\end{equation}
It is clear that $T(\bm{X},\bm{Y};\theta) \leq \gamma_{(\theta, \eta^T)}\{|W(\bm{X},\bm{Y};\theta)|\}$, i.e., $T(\bm{X},\bm{Y};\theta)$ is dominated by $\gamma_{(\theta, \eta^T)}\{|W(\bm{X},\bm{Y};\theta)|\}$.
Since $X_{i} = \sum_{j=1}^{n_i} I\{U_{ij} \leq $ $e^{(\eta_{i}+\theta)/2}\}$ and 
$Y_{i} = \sum_{j=1}^{m_i} I\{V_{ij} \leq$ $e^{(\eta_{k}-\theta)/2}\}$, the mapping  $T(\bm{X},\bm{Y};\theta)$ is a function of ${\bm U} = \{U_{ij}, 1 \leq j \leq n_i, 1 \leq i \leq K\}$, ${\bm V} = \{V_{ij}, 1 \leq j \leq m_i, 1 \leq i \leq K\}$ and $(\theta, \bm{\eta}^T).$
Thus, for a given $\theta$, the distribution of $T(\bm{X},\bm{Y};\theta)$ still depends on the nuisance parameter $\bm{\eta}$. However, we always have 
\begin{equation}\label{eq:T}
    \mathbf{P}\left\{ T(\bm{X},\bm{Y};\theta) \leq \alpha \right\} \geq \mathbf{P}\left[ \gamma_{(\theta, \eta^T)}\{|W(\bm{X},\bm{Y};\theta)|\} \leq \alpha \right] = \alpha. 
\end{equation}
Thus, a Borel set corresponding to (\ref{eq:B}) is $B_\alpha = (0, \alpha]$ which is free of both $\theta$ and $\bm \eta$.   

Following (\ref{eq:G0}), the level-$\alpha$ repro samples confidence set for $\theta$ is: 
\begin{align}
    \Gamma_\alpha(\bm{x}_{obs}, \bm{y}_{obs}) & = \big\{\theta: \exists \, ({\bm u}^*, {\bm v}^*) \hbox{ and } {\bm \eta} \hbox{ such that }   ({\bm x}^{obs}, {\bm y}^{obs}) = ({\bm x}^*, {\bm y}^*),  \nonumber 
  \\
    & \qquad\qquad\qquad 
    T(\bm{x}^*, \bm{y}^*;\theta) \leq \alpha
    \big\} \nonumber \\
    & = \big\{\theta: \exists \, ({\bm u}^*, {\bm v}^*) \hbox{ and } {\bm \eta} \hbox{ such that }   ({\bm x}^{obs}, {\bm y}^{obs}) = ({\bm x}^*, {\bm y}^*),  \nonumber 
  \\
    & \qquad\qquad\qquad 
    T(\bm{x}^{obs}, \bm{y}^{obs};\theta) \leq \alpha
    \big\} \nonumber \\
    & = \left\{\theta: 
    T(\bm{x}_{obs}, \bm{y}_{obs};\theta) \leq \alpha
    \right\},  \label{eq:Gamma}
\end{align}
where ${\bm x}^* = (x_1^*, \ldots, x_K^*)^T$ and ${\bm y}^* = (y_1^*, \ldots, y_K^*)^T$  with 
$x_i^* = \sum_{j = 1}^{n_i} I\{ u^*_{ij} \leq e^{(\theta + \eta_{i})/2} \}$ and $y_i^* = \sum_{j = 1}^{m_i} I\{v^*_{ij} \leq e^{(\theta - \eta_{i})/2} \},$ for $i = 1 \leq i \leq K.$
The first equation of (\ref{eq:Gamma}) follows the repro samples approach. The last equation holds since, for a given $\theta$, there always exist $({\bm u}^*, {\bm v}^*)$ and $\bm \eta$ such that $({\bm x}^{obs}, {\bm y}^{obs}) = ({\bm x}^*, {\bm y}^*)$.  

By equation $(\ref{eq:T})$, we 
have the following theorem that  $\Gamma_\alpha(\bm{x}_{obs}, \bm{y}_{obs})$ in (\ref{eq:Gamma}) is a level-$\alpha$ confidence set for the common log odds ratio $\theta^{(o)}$. 
\begin{theorem} 
Under the above setup and suppose the random sample are generated using the parameter values $(\theta^{(o)}, \bm{\eta}^{(o)T})$, i.e.,
$X_{i} = \sum_{j=1}^{n_i} I{\{U_{ij} \leq e^{(\eta_{i}^{(o)}+\theta^{(o)})/2}\}}$ and 
$Y_{i} = \sum_{j=1}^{m_i} I{\{V_{ij} \leq e^{(\eta_{i}^{(o)}-\theta^{(o)})/2}\}}$, we have
$$
\mathbf{P}\left\{\theta^{(o)} \in \Gamma_\alpha(\bm{X}, \bm{Y}) \right\} \ge \alpha. 
$$
\end{theorem}

\subsection{Monte-Carlo implementation and computing algorithm} 

To construct the level-$\alpha$ confidence set in (\ref{eq:Gamma}), we need to calculate $T(\bm{x}_{obs}, \bm{y}_{obs};\theta) = \min_{\widetilde \eta} \gamma_{(\theta, \widetilde \eta^T)}\{|W(\bm{x}_{obs},\bm{y}_{obs};\theta)|\}$, for a potential $\theta$ value. This can be done by using a Monte-Carlo method to approximate  $\gamma_{(\theta, \widetilde \eta^T)}\{|W(\bm{x}_{obs},\bm{y}_{obs};\theta)|\}$. Specifically, 
for any set of fixed $(\theta, \widetilde {\bm{\eta}}^T)$, we can approximate the function $\gamma_{(\theta, \widetilde \eta^T)}\{t\}$ by 
\begin{equation} \label{eq:p}
    \gamma_{(\theta, \widetilde \eta^T)}\{t\} \approx \frac1M \sum_{s=1}^M {I}\left\{ \big|W(\widetilde{\bm{x}}^{(s)},\widetilde{\bm{y}}^{(s)};\theta)\big| < t
\right\}, 
\end{equation}
where 
$\widetilde {\bm x}^{(s)} = (\widetilde x_{1}^{(s)}, \ldots, \widetilde x_{K}^{(s)})^T$, 
$\widetilde {\bm y}^{(s)} = (\widetilde y_{1}^{(s)}, \ldots, \widetilde y_{K}^{(s)})^T$, 
$\widetilde x_{i}^{(s)} = \sum_{i=1}^{n_i} I{\{U_{ij}^{(s)} \leq e^{(\widetilde \eta_{i}+\theta)/2}\}}$, $\widetilde y_{i}^{(s)} = \sum_{j=1}^{m_i} I{\{V_{ij}^{(s)} \leq e^{(\widetilde \eta_{i}-\theta)/2}\}}$ and $(U_{ij}^{(s)}, V_{ij}^{(s)})$ are simulated iid $U(0,1)$ random numbers, for $s = 1 \ldots, M$. Thus, we can approximate  $\gamma_{(\theta, \widetilde {\eta}^T)}\{|W(\bm{x}_{obs},\bm{y}_{obs};\theta)|\}$, 
which is only a function of $(\theta, \widetilde {\bm \eta}^T)$. We then call an optimization program to find its minimum value over $\tilde {\bm \eta}$, and it leads to $T(\bm{x}_{obs},\bm{y}_{obs};\theta)$ that is a 
function of $\theta$ when given $(\bm{x}_{obs},\bm{y}_{obs})$. 

We provide below a computing algorithm: 

\begin{breakablealgorithm}
\caption{Calculation of confidence interval of common log odds ratio}
\begin{itemize}
  \setlength{\itemindent}{.5 in}
  \item[Step 1:]
  Compute $W_{MH}(\bm{x}_{obs},\bm{y}_{obs})$ and select grids for $\theta$ on its range, say $\theta_{1}, \cdots,\theta_{Q}$.  
		
  \item[Step 2:]
  Set $\widetilde{\Theta}=\emptyset$. For $m=1,2,\cdots,Q$, repeat the following computation:
  \begin{itemize}
    \setlength{\itemindent}{.6in}
      \item[Step 2a:]
      Calculate 
     {\small \begin{align*}
      T(\bm{x}_{obs}, \bm{y}_{obs};\theta_{m}) 
      = \min_{\widetilde \eta}
      \frac1M \sum_{s=1}^M {I}\left\{ \big|W(\widetilde{\bm{x}}^{(s)},\widetilde{\bm{y}}^{(s)};\theta_m)\big| < |W_{MH}(\bm{x}_{obs},\bm{y}_{obs}) - \theta_{m}|
      \right\},
      \end{align*}}
      where $\widetilde{\bm{x}}^{(s)} = (\widetilde x_{1}^{(s)}, \ldots, \widetilde x_{K}^{(s)})^T$, 
$\widetilde {\bm{y}}^{(s)} = (\widetilde y_{1}^{(s)}, \ldots, \widetilde y_{K}^{(s)})^T$, 
$\widetilde x_{i}^{(s)} = \sum_{i=1}^{n_i} I\{u_{ij}^{(s)} \leq $ $ e^{(\widetilde \eta_{i}+\theta_{m})/2}\}$, $\widetilde y_{i}^{(s)} = \sum_{j=1}^{m_i} I\{v_{ij}^{(s)} \leq $ $e^{(\widetilde \eta_{i}-\theta_{m})/2}\}$ and $(u_{ij}^{(s)}, v_{ij}^{(s)})$ are simulated iid $U(0,1)$ random numbers, for $s = 1 \ldots, M$.

	\item[Step 2b:]
	For given $0<\alpha<1$, if $T(\bm{x}_{obs}, \bm{y}_{obs};\theta_{m}) \leq \alpha$, update $\widetilde{\Theta}=\widetilde{\Theta}\cup\theta_{m}$.	
    \end{itemize}
		
\item[Step 3:]
Compute $\min\{\widetilde{\Theta}\} $ and $\max \{\widetilde{\Theta}\} $. The $100\alpha\%$ confidence interval for $\theta$ is $[\min \{\widetilde{\Theta}\} ,\max \{\widetilde{\Theta}\} ]$.
\end{itemize}
\end{breakablealgorithm}

\section{Simulation Studies}

In this section, we examine the empirical performance of our repro samples method on making inference for the common log odds ratio $\theta$, and also make comparisons with the popular Mantel-Haenszel and Peto methods. In particular, we compare the empirical coverage probabilities  
and average lengths of the confidence intervals based on 500 replications with M =1000. 

To generate simulated data, we design a context similar to the structure of Avandia dataset, following \citet{Tianetal2009} and \citet{LiuLiuXie2014}. Concretely, $K=48$ independent $2\times 2$ tables are generated using the same sample sizes of Avandia dataset. The incidence rate $\pi_{0i}^{(o)}$ in $i$th trial is generated from a uniform distribution $U(0,0.08)$. Then the incidence rate $\pi_{1i}^{(o)}$ is determined by relationship ${\rm logit}(\pi_{1i}^{(o)})=\theta^{(o)}+{\rm logit}(\pi_{0i}^{(o)})$, where several true common log odds ratio values $\theta^{(o)}$ under various scenarios are examined. Finally, the $i$th table is simulated by the binomial distributions with the generated $(\pi_{0i}^{(o)},\pi_{1i}^{(o)})$.

In the implementation of our repro samples algorithm, we confine our potential $\theta$ values within the 99.95\% confidence interval of the true $\theta^{(o)}$ obtained using the Mantel-Haenszel approach. For each $\theta$, it is noted that the nuclear mapping involves the minimization over $\bm{\eta}=(\eta_{1},\cdots,\eta_{K})^T$ with $K=48$. We apply the R function `optim' in the package `stats' to find the minimum value. In the implementation of minimization via `optim', an initial value of $\bm{\eta}$ need to be specified. Recall that $\eta_{i}=\log\big(\pi_{1i}/(1-\pi_{1i})\big)+\log\big(\pi_{0i}/(1-\pi_{0i})\big)$ for $k=1,\cdots,K$. Then, if $i$th trial has nonzero events in both groups, the initial value of $\eta_{k}$ is given by
$\hat{\eta}_{i}=\log\Big(\frac{\hat{\pi}_{1i}}{1-\hat{\pi}_{1i}}\Big)+\log\Big(\frac{\hat{\pi}_{0i}}{1-\hat{\pi}_{0i}}\Big)$,  
where $\hat{\pi}_{1i}=x_i/n_i$ and $\hat{\pi}_{0i}=y_i/m_i$. However, it will not work for trials with zero events in one arms. In view of the similarity among all the trials, we use $\min\{\hat{\eta}_{k}:k\mbox{th trial has nonzero events in both groups}, 1\leq k \leq K\}$ as the initial value of $\eta_{k}$ for trials with zero events in one or both group.

Tables 2 to 4 list the empirical results based on $500$ data replications when the common odds ratio $\theta$ takes different values. 
Based on these tables, we can see that the proposed repro samples method produces valid confidence intervals for the prespecified confidence level of 95\% for all different $\theta$ values. 
The empirical coervages of the Mantel-Haenszel method are mostly on target, although a few of them have slightly undercoverage rates. 
Peto method only works for moderate $\theta$'s, and breaks down for those large and small $\theta$'s. 
In addition, we can see that interval lengths of repro samples are similar but slightly longer than those obtained using Mantel-Haenszel method. To ensure the coverage rates across all cases,  the repro samples approach is slightly conservative, which is expected by equation (\ref{eq:T}).

Finally, we conduct a numerical study to demonstrate that our proposed repro samples method
can effectively extract information hidden in the zero-total-event studies for the  common odds ratio parameter. 
Suppose we have two datasets, both of which include two non-zero-total-event studies and three zero-total-event studies: (a)  (3/100, 2/100), (2/300, 1/300), (0/600, 0/300), (0/600, 0/300), (0/300, 0/300); and (b) (2/100, 2/100), (1/50, 1/50), (0/100, 0/300), (0/100, 0/300), (0/100, 0/300). For each of the two datasets, we use
our algorithm to obtain 
the two level-$95\%$ confidence intervals for the common log odds ratio $\theta^{(o)}$, one using all five studies and the other using only the two non-zero-total-event studies (excluding the three zero-total-event studies). 
Figure 1 depicts the comparisons of these two sets of intervals. Based on the figure, we can see that the confidence intervals obtained by excluding the three zero-total-event studies are significantly wider than the intervals obtained by including them. This set of results further affirms the conclusion that zero-total-event studies has information and impacts the inference of the common odds ratio as discussed in \cite{Xieetal2018}. Overall, our repro samples method provides a solution to effectively include zero-total-event studies in the analysis of the common odds ratio parameter in meta-analysis.

\begin{table}
	\caption{Comparisons of MH, Peto and repro samples by mimicking the structure of Avandia dataset with true common odds ratio $\theta^{(o)}$ being 1.0 to 1.9.}
	\vspace{-2mm}
	\centering
	\footnotesize{
		\begin{tabular*}{\textwidth}{cccccccccccccccccccccc}
			\hline \hline
			&&\multicolumn{10}{c}{True odds ratio}\\
			\cline{3-12}
			&&1.0&1.1&1.2&1.3&1.4&1.5&1.6&1.7&1.8&1.9
			\\
			\hline
			MH&CP&0.946&0.944&0.940&0.936&0.952&0.960&0.962&0.956&0.954&0.964
			\\
			&Length& 0.772&0.753&0.743&0.730&0.721&0.720&0.705&0.700&0.691&0.687
			\\

			Peto&CP&0.946&0.944&0.940&0.938&0.956&0.968&0.966&0.960&0.958&0.966
			\\			&Length&0.769&0.747&0.729&0.710&0.696&0.684&0.666&0.655&0.638&0.630
			\\
  Repro&CP&0.974&0.966&0.974&0.962&0.966&0.970&0.976&0.970&0.964&0.980
			\\											&Length&0.891&0.870&0.858&0.834&0.830&0.829&0.818&0.804&0.801&0.792 
			\\
			\hline \hline
			\\
		\end{tabular*}
	}
\end{table}

\begin{table}
	\caption{Comparisons of MH, Peto and repro samples by mimicking the structure of Avandia dataset with true common odds ratio $\theta^{(o)}$ being 2 to 9.}
	\vspace{-2mm}
	\centering
	\footnotesize{
		\begin{tabular*}{\textwidth}{cccccccccccccccccccccc}
			\hline \hline
			&&&&\multicolumn{8}{c}{True odds ratio}\\
			\cline{5-12}
			&&&&2&3&4&5&6&7&8&9
			\\
			\hline			
                &&MH&CP& 0.958&0.956&0.952&0.952&0.946&0.934&0.944&0.950
			\\			&&&Length&0.681&0.643&0.637&0.623&0.616&0.609&0.605&0.599
		    \\		               
			&&Peto&CP&0.970&0.840&0.466&0.066&0&0&0&0
			\\
			&&&Length&0.618&0.531&0.478&0.457&-&-&-&-
			\\
&&Repro&CP&0.974&0.976&0.976&0.974&0.956&0.960&0.978&0.966
			\\			&&&Length&0.792&0.745&0.736&0.724&0.715&0.705&0.701&0.695
			\\
			\hline \hline
			\\
		\end{tabular*}
	}
\end{table}

\begin{table}
	\caption{Comparisons of MH, Peto, and repro samples by mimicking the structure of Avandia dataset with true common odds ratio $\theta^{(o)}$ being 1/1.8 to 1.8.}
	\vspace{-2mm}
	\centering
	\footnotesize{
		\begin{tabular*}{\textwidth}{cccccccccccccccccccccc}
			\hline \hline
			&&&\multicolumn{9}{c}{True odds ratio}\\
			\cline{4-12}
			&&&1/1.8&1/1.6&1/1.4&1/1.2&1&1.2&1.4&1.6&1.8
			\\
			\hline			&MH&CP&0.972&0.952&0.952&0.950&0.946&0.940&0.952&0.962&0.954
			\\
			 &&Length&0.897&0.864&0.832&0.804&0.772&0.743&0.721&0.705&0.691
			\\
			&Peto&CP&0.972&0.950&0.956&0.954&0.946&0.940&0.956&0.966&0.958
			\\
			&&Length&0.885&0.861&0.832&0.805&0.769&0.729&0.696&0.666&0.638
			\\
			&Repro&CP&0.978&0.958&0.966&0.972&0.974&0.974&0.966&0.976&0.964
			\\
			&&Length&1.016&0.987&0.954&0.923&0.891&0.858&0.830&0.818&0.801
			\\
			\hline \hline
			\\
		\end{tabular*}
	}
\end{table}

\begin{figure}
  \centering
  \subfloat[]{\includegraphics[width=3.2in]{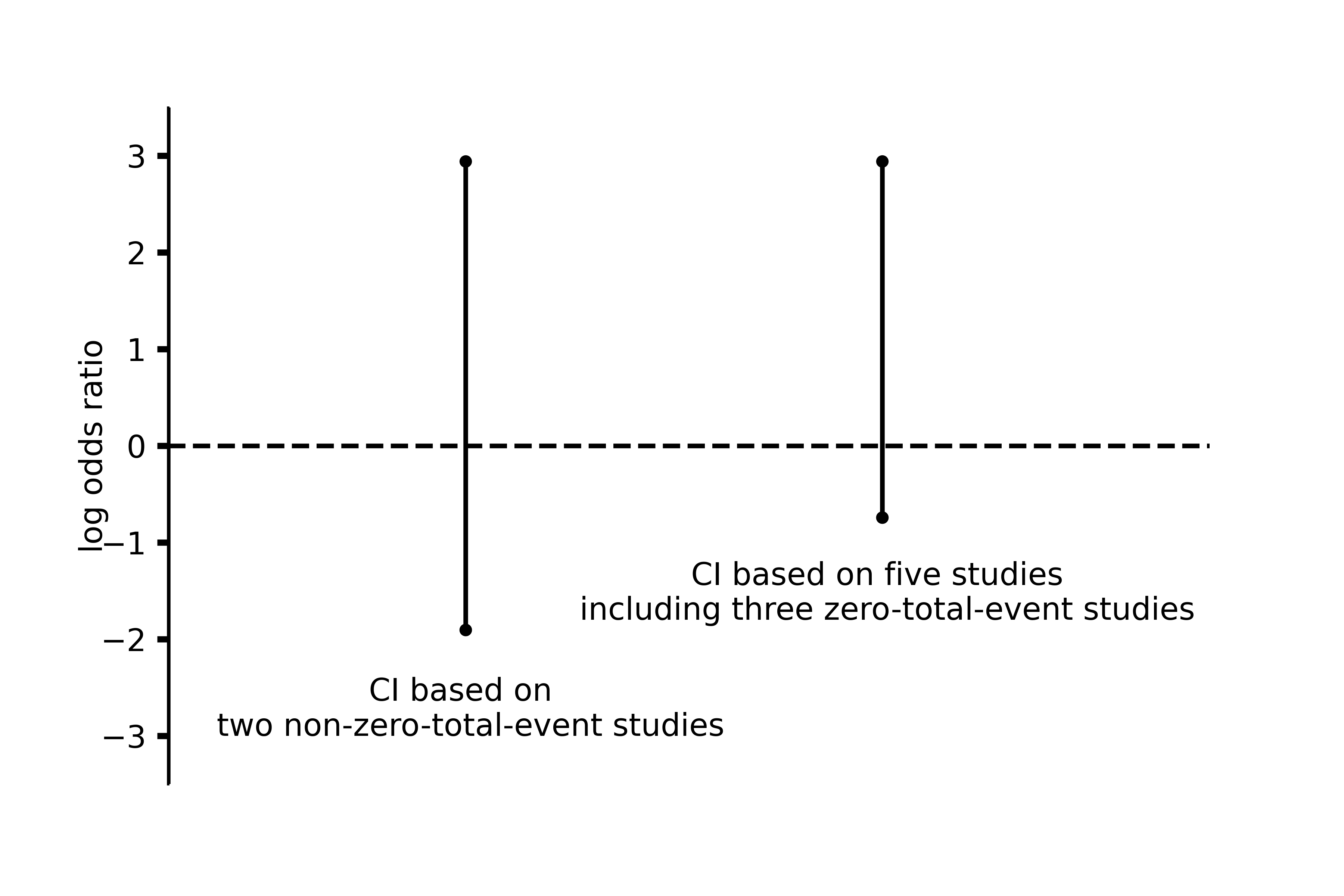}}
  \subfloat[]{\includegraphics[width=3.2in]{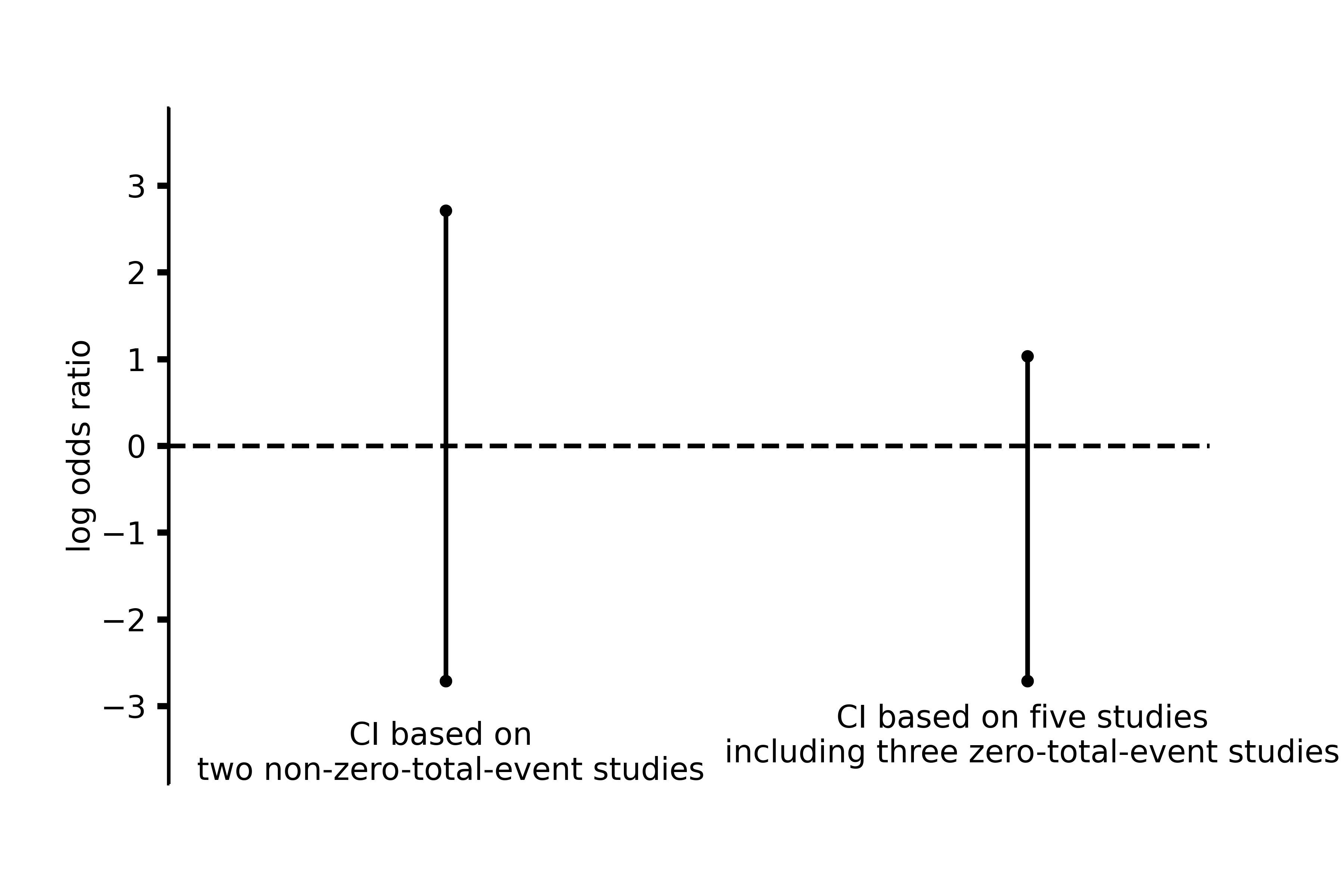}}
  \caption{Illustration of the impact of zero-total-event studies on level-$95\%$ confidence intervals of common log odds ratio: analysis using two studies (removing zero-total-event studies) versus analysis using all five studies. The two datasets used are: (a) $(3/100,2/100),$ $ (2/300,1/300), $ $(0/600,0/300),$ $(0/600,0/300),$ $ (0/300,0/300)$; and (b) $(2/100,2/100),$ $(1/50,1/50),$ $(0/100,0/300),(0/100,0/300),$ $(0/100,0/300)$.}
\end{figure}

\section{Real Data Analysis}

Avandia dataset \citep{NissenWolski2007} includes data from $K =48$ independent clinical trials to examine its effect on cardiovascular morbidity and mortality. In fact, Avandia is the trade name of drug rosiglitazone, which is widely used for treatment of type 2 diabetes mellitus. Among the 48 trials, there are 46 small trials with sample size at most 1172 in one arm and 2 large trials with sample sizes at least 1456 in one group. The two large trials are called Diabetes Reduction
Assessment with Ramipiril and Rosiglitazone
Medication (DREAM) and A Diabetes Outcome Prevention Trial (ADOPT), respectively. In this dataset, the events of myocardial infarction and cardiovascular death have very low incidence rate. Thus, many trials do not contain any or only contain very few interested events, especially for death from cardiovascula causes. Specially, there exist many trials with zero events in one of two arms and zero-total-event trials. Among the 48 trials, 10 reports no events for myocardial infarction and 25 reports no events for cardiovascular death in both of treatment and control groups. The entire dataset could be found in Table I of the supplementary material of \citet{Tianetal2009}. It is an extremely non-trial and challenging task to effectively incorporate these studies in a meta-analysis
\citep{FinkelsteinLevin2012,Xieetal2018}. In \cite{Xieetal2018}, the authors made a definite conclusion that zero-total-event trials have information of the common odds ratio. 
Here, 
we apply our newly developed finite sample method along with the widely used Mantel-Haenszel and Peto methods to construct confidence intervals for the common odds ratio. 

\begin{table}
	\caption{Analysis of Avandia dataset: $95\%$ confidence intervals of common odds~ratio }
	\vspace{-1mm}
	\centering
	\small{
		\begin{threeparttable} 
			\begin{tabular*}{\textwidth}{cccccccccccc}		
				\hline \hline
				&&&&&&&&&MI&CVD
				\\
				\hline
				&&&&&&&&MH&(1.029,1.978)&(0.984,2.930)
				\\
				&&&&&&&&Peto&(1.031,1.979)&(0.980,2.744)
				\\
				&&&&&&&&Repro-1&(0.982,2.118)&(0.962,3.283) 
                \\
                &&&&&&&&Repro-2&(0.962,2.165)&(0.846,3.802)
				\\
				\hline \hline
				\\
			\end{tabular*}
			\vspace{-0.8cm}
			\begin{tablenotes}
				\item MI is for  myocardial infarction; CVD is for cardiovascular death. Repro-1 uses data from all the 48 trials; Repro-2 excludes zero-total-event trials.
			\end{tablenotes}
		\end{threeparttable}
	}
\end{table}
The 95\% confidence intervals for common odds ratios of myocardial infarction and cardiovascular death obtained by these three approaches, denoted as MH, Peto, Repro-1, respectively, are listed in Table 5. For the endpoint of cardiovascular death, three methods output the similar results. Three confidence intervals all include the value of 1. Thus, all of them suggest that the drug rosiglitazone has no statistically significant effect on mortality of cardiovascular death. Our repro samples method, however, obtained smaller lower end of confidence interval and show stronger evidence that the drug rosiglitazone has no statistically significant effect on mortality of cardiovascular death. 

As for myocardial infarction, the results are quite different. The confidence intervals of conventional Mantel-Haenszel and Peto methods exclude the value 1, while that using the repro samples method includes it. 
According to Mantel-Haenszel and Peto means, the drug rosiglitazone has statistically significant effect. 
However, using the repro samples method, we could not conclude that the drug rosiglitazone has a statistically significant effect on myocardial infarction.

Finally, we examine the impact of zero-total-event studies on the confidence intervals of common log odds ratio in the Avandia dataset. Specifically, we re-run our repro sample algorithm by deleting the zero-total-event studies, and compare the confidence intervals obtained without including zero-total-event studies, denoted by Repro-2 in Table 5, with those previously obtained including these zero-total-event studies. For the event of myocardial infarction, there are $10$ zero-total-event studies. For the event of cardiovascular death there are $25$ zero-total-event studies. 
From Table 5, we can see that intervals with and without including the zero-total-event studies are quite different. The intervals with zero-total-event studies are narrower than those without including zero-total-event studies. This shows that utilizing zero-total-event studies in meta-analysis is important and beneficial for the inference of the common log odds ratio in general. It reaffirms our conclusion that the zero-total-event studies has information and impacts the inference of the common odds~ratio.

\section{Discussion}

Questions on whether a zero-total-event study contains any information for the common odds ratio in meta-analysis of $2\times2$ tables and how to incorporate such studies when making inference for the common odds ratio have long been debated and remain to be open in statistics \citep[cf., ][]{FinkelsteinLevin2012, Xieetal2018}. The difficulty is due to the lack of mathematical definition for $0/0$ and also because most meta-analysis approaches rely on normality and large sample theories both of which do not apply for the zero-total-event studies. 
In this article and by using the recent developed repro samples inferential framework, we are able to develop a finite-sample approach to make inference for the common odds ratio. The developed inference procedure has guaranteed theoretical performance and is validated in numerical studies. It provides an affirmative answer to the set of open research questions. 

The repro sample framework is developed based on the ideas of  inversion,  matching of artificial and observed samples, and simplifying  uncertainty quantification through a Borel set concerning $U$. It does not need any regularity conditions, nor relies on any large sample theories. It can provide finite sample inference with few assumptions, and is an ideal tool to address some difficult and complicated inference problems. In this article, 
we have used it to develop a novel approach to answer the unresolved questions concerning the use of zero-total-event studies in meta-analysis. The repro samples method can also be used to develop new finite-sample procedures in other meta-analysis settings; for instance, developing a new finite-sample approach to perform meta-analysis and combine information in a random-effects model with only a few studies, a setting studied in \cite{michael2019exact}.   
Furthermore, the repro samples method is also very effective for other irregular inference problems that involve discrete or non-numerical parameters. For instance, \cite{XieWang2022} and \cite{WangXieZhang2022} provided solutions for two highly nontrivial problems in statistics: a) how to quantify the uncertainty in the estimation of the unknown number of components and make inference for the associated parameters in a Gaussian mixture; b) how to quantify the uncertainty in model estimation and construct confidence sets for the unknown true model, the regression coefficients, or both true model and coefficients jointly in high dimensional regression models. We anticipate these developments will stimulate further developments to address more complicated and non-trivial inference problems in statistics and data science where a solution is currently unavailable or cannot be easily obtained.

\section{Acknowledgment}

Xie's research is supported in part by NSF grants DMS2015373, DMS2027855, DMS2311064 and DMS-2319260. Chen's research is supported partly by Humanity and Social Science Research Foundation of Ministry of Education (MOE) of China (21YJA910002).

\bibliographystyle{asa}
\bibliography{OddsRatioReproSamples}

\end{document}